\documentclass[12pt]{article}

\usepackage{scicite}

\usepackage{times}

\usepackage{amsmath}
\usepackage{amsfonts}
\usepackage{amssymb}
\usepackage{graphicx}

\usepackage{color}

\newcommand{\captiontitle}[1]{{\bf #1}}

\newenvironment{sciabstract}{%
\begin{quote} \bf}
{\end{quote}}

\title{Realization of nearly dispersionless bands with strong orbital anisotropy from destructive interference in twisted bilayer MoS$_2$}

\author{Lede Xian,${}^{1}$ Martin Claassen,${}^{2}$ Dominik Kiese,${}^{3}$ Michael M. Scherer,${}^{3}$ \\
Simon Trebst,${}^{3}$ Dante M. Kennes,${}^{1,4\ast}$ Angel Rubio${}^{1,2,5\ast}$\\
\\
\normalsize{${}^{1}$Max Planck Institute for the Structure and Dynamics of Matter and} \\
\normalsize{Center for Free Electron Laser Science, Luruper Chaussee 149, 22761 Hamburg, Germany}\\
\normalsize{${}^{2}$Center for Computational Quantum Physics,}\\
\normalsize{Simons Foundation Flatiron Institute, New York, NY 10010 USA}\\
\normalsize{${}^{3}$Institute for Theoretical Physics, University of Cologne, 50937 Cologne, Germany}\\
\normalsize{${}^{4}$Institut f\"ur Theorie der Statistischen Physik, RWTH Aachen University and }\\
\normalsize{JARA-Fundamentals of Information Technology, 52056 Aachen, Germany }\\
\normalsize{${}^{5}$Nano-Bio Spectroscopy Group,  Departamento de Fisica de Materiales,}\\
\normalsize{Universidad del Pa\'is Vasco, UPV/EHU- 20018 San Sebasti\'an, Spain}\\
\\
\normalsize{$^\ast$To whom correspondence should be addressed; }\\
\normalsize{E-mail:  dante.kennes@rwth-aachen.de; angel.rubio@mpsd.mpg.de}
}

\date{}

\begin{document} 


\baselineskip24pt


\maketitle



\begin{sciabstract}
 Recently, the twist angle between adjacent sheets of stacked van der Waals materials emerged as a new knob to engineer correlated states of matter in two-dimensional heterostructures in a controlled manner, giving rise to emergent phenomena such as superconductivity or correlated insulating states. Here, we use an {{\slshape \textbf{ab initio}}} based approach to characterize the electronic properties of twisted bilayer MoS$_2$. We report that, in marked contrast to twisted bilayer graphene, slightly hole-doped MoS$_2$ realizes a strongly asymmetric p$_{\rm x}$-p$_{\rm y}$ Hubbard model on the honeycomb lattice, with two almost entirely dispersionless bands emerging due to destructive interference. We study the collective behavior of twisted bilayer MoS$_2$ in the presence of interactions, and characterize an array of different magnetic and orbitally-ordered correlated phases, which may be susceptible to quantum fluctuations giving rise to exotic, purely quantum, states of matter.

\end{sciabstract}

\section*{Introduction}

Two-dimensional van der Waals materials constitute a versatile platform to realize quantum states by design, as they can be synthesized in many different stacking conditions \cite{liu2020disassembling}, offer a wide variety of chemical compositions and are easily manipulated by back gates, strain and the like. Stacking two sheets of van der Waals materials atop each other at a relative twist has recently emerged as a vibrant research direction to enhance the role of electronic interactions, with first reports on  twisted bilayer graphene \cite{cao2018a,cao2018b,Yankowitz18,Kerelsky18,Choi19} and other van der Waals materials stacked atop each other at a twist  \cite{Jin18,tran2019evidence,alexeev2019resonantly,seyler2019signatures,Xian18,Naik18,Wang19TMD,Scuri19,Kerelsky19,Andersen19,Kennes2020} displaying features of correlated physics that afford an unprecedented level of control. In particular, bi-, tri- and quadruple-layer graphene as well as twisted few-layer transition metal dichalcogenides (TMDs) are currently under intense experimental scrutiny \cite{lu2019superconductors,stepanov2019interplay,arora2020superconductivity,chen2020nature,liu2019spin,shen2019observation,cao2019electric,he2020,tutuc2019,Wang19TMD}.  By forming a Moir\'e supercell at small twist angles, a large unit cell in real space emerges for twisted systems, which due to quantum interference effects leads to a quasi-two-dimensional system with strongly quenched kinetic energy scales. This reduction in kinetic energy scale, signaled by the emergence of flat electron bands, in turn enhances the role of electronic interactions in these systems.

Whereas the flatting of band dispersions in two-dimensional Moir\'e superlattices results mainly from the localization of charge density distributions by the Moir\'e potential, a well-known alternate pathway to flat bands can occur in certain lattices such as the Lieb and the Kagome lattices, where geometric considerations permit the formation of perfectly-localized electronic states on plaquettes and hexagons, respectively, that are eigenstates of the kinetic Hamiltonian due to destructive interference between lattice hopping matrix elements \cite{liu_2014}. Such flat band systems can give rise to many interesting phenomena, such as the formation of nontrivial topology when time reversal symmetry is broken, or other exotic quantum phases of matter due to their susceptibility to quantum fluctuations and electronic correlations \cite{leykam2018artificial}.
        
Here, we demonstrate that both flat-band mechanisms can be engineered to coexist in twisted bilayers of MoS$_2$ (tbMoS$_2$); a TMD of direct experimental relevance that has been extensively studied from synthesis to applications \cite{mak2010,wang2018synthesis}.
We confirm that families of flat bands emerge when two sheets of MoS$_2$ are stacked at a twist \cite{Naik18,naik2019origin} due to Moir\'e potentials. Our large-scale \textit{ab initio} based simulations show that while the first set of engineered flat bands closest to the edge of the band gap with twist angles close to $\Theta \approx 0^{\circ}$ can be used to effectively engineer a non-degenerate electronic flat band in analogy to a single layer of graphene at meV energy scales, more intriguingly, the next set of flat bands instead realizes a strongly asymmetric flat band p$_{\rm x}$-p$_{\rm y}$  honeycomb lattice \cite{wu2007prl,wu2008prb}. Both of these families of bands should be accessible experimentally via gating.  The strongly asymmetric nature of this p$_{\rm x}$-p$_{\rm y}$ honeycomb lattice is in marked contrast to the much discussed case of twisted bilayer graphene, where an approximately symmetric version of such a Hamiltonian is now believed to describe the low-energy flat band structures found at small twist angle \cite{bistritzer2011,PhysRevB.98.045103,PhysRevB.99.195455,PhysRevResearch.1.033072,TBGFRG}. The strongly asymmetric p$_{\rm x}$-p$_{\rm y}$ honeycomb model realized here features two almost entirely dispersionless flat bands that touch the top and the bottom of graphene-like Dirac bands at the Gamma point, respectively. These flat bands originate from destructive interference, in analogy to flat bands in the Lieb and the Kagome lattices \cite{liu_2014}, and will be referred to as ultra-flat bands in the following discussion, to distinguish them from other bands with quenched kinetic energy scales. In addition, these ultra-flat bands are topologically non-trivial once time-reversal symmetry is broken via spin-orbital coupling \cite{zhang2014}.  Previously, the p$_{\rm x}$-p$_{\rm y}$ model was studied in the context of cold gases where exotic correlated phases were predicted \cite{wu2007prl,lee2010fwave,wu2010f}, as well as in semiconductor microcavities \cite{jacqmin2014} and certain 2D organometallic frameworks \cite{liu2013omf,su2018}. Our findings elevate tbMoS$_2$ to a novel platform where effects of ultra-flat bands can be studied systematically in a strongly-correlated solid-state setting.
%
Notably, in the strong-coupling regime, the  p$_{\rm x}$-p$_{\rm y}$ model amended by Hubbard and Hund's interactions gives rise to a spin-orbital honeycomb model which -- depending on the specific parameters and symmetries of the model -- hosts magnetic, orbital as well as valence-bond orderings, or even more exotic quantum spin-orbital liquid phases~\cite{PhysRevB.90.094422,PhysRevB.98.245103,PhysRevB.100.205131}.

With this, our work adds an unprecedented type of lattice model -- the 
highly asymmetric p$_{\rm x}$-p$_{\rm y}$ Hubbard model -- to the growing list of systems that can effectively be engineered using the twist angle between multiple layers. This is particularly intriguing as we maintain the full advantages that come with two-dimensional van der Waals materials, such as relative simplicity of the chemical composition and controllability of the material properties; e.g. of the filling (by a back gate), electric tunability (by displacement fields) or the band width of the model (by the twist angle).

\section*{Results}

\subsection*{{{\slshape \textbf{ab initio}}} characterization of twisted MoS$_2$}
\begin{figure}
    \centering
	 \includegraphics[width = \columnwidth]{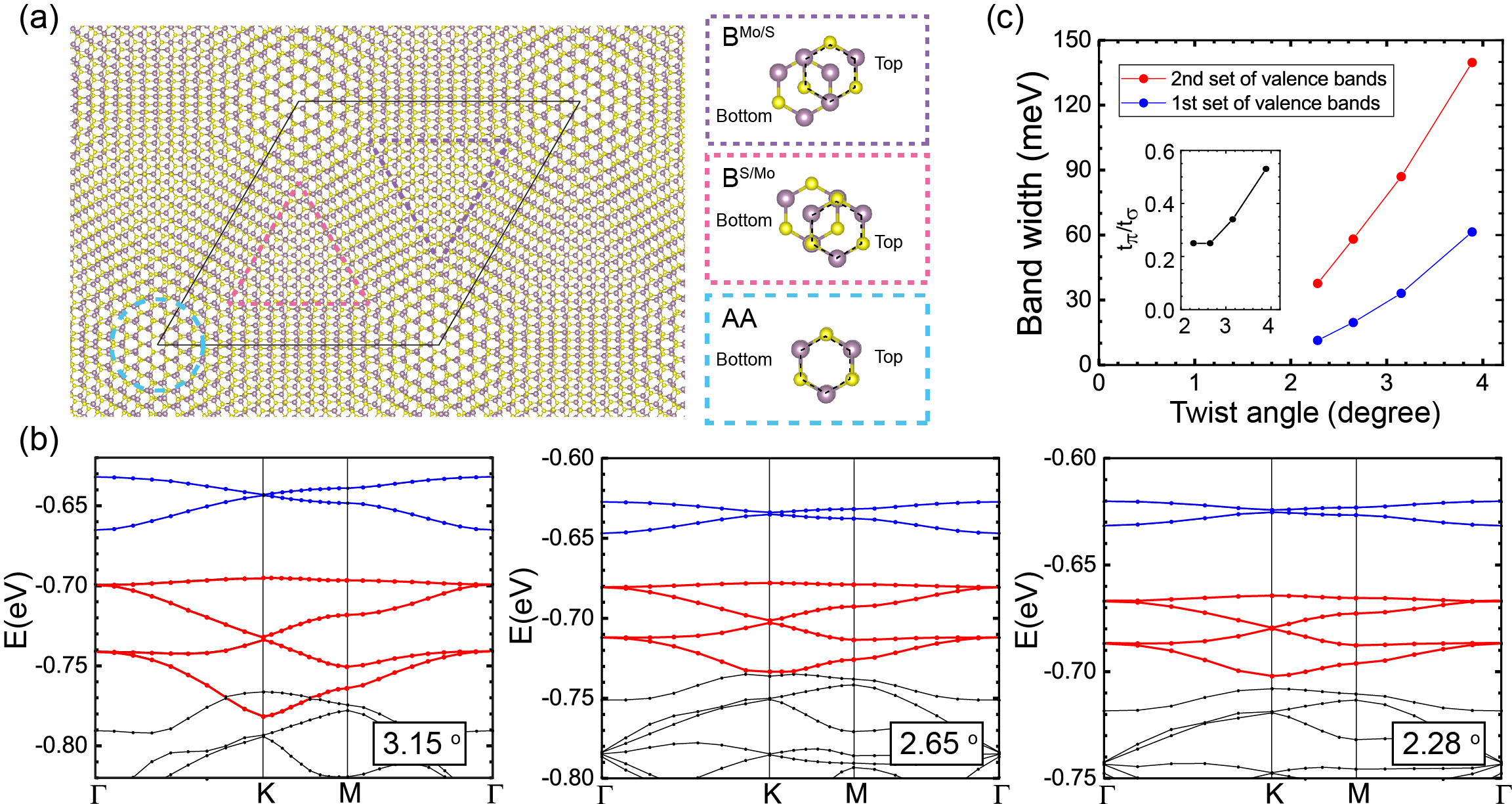}
	  \caption{\captiontitle{Atomic and electronic structures of twisted bilayer MoS$_2$.} (a) Atomic structure of tbMoS$_2$ at $\Theta=3.15^\circ$. Local atomic arrangements of the three different regions in the Moir\'e unit cell are indicated in the right panels  (b) Evolution of low-energy band structures at the top of the valence bands of tbMoS$_2$ with decreasing small twist angles. The first set and the second set of valence bands are highlighted with blue and red lines, respectively.  (c) Evolution of the band width of the first set and the second set of valence bands with decreasing twist angles.    Inset: twist angle dependence of the ratio of the hopping amplitudes $t_\pi$ and $t_\sigma$ in the p$_{\rm x}$-p$_{\rm y}$ honeycomb lattice.      }
    \label{fig:fig1}
\end{figure}

\begin{figure}
    \centering
	 \includegraphics[width = \columnwidth]{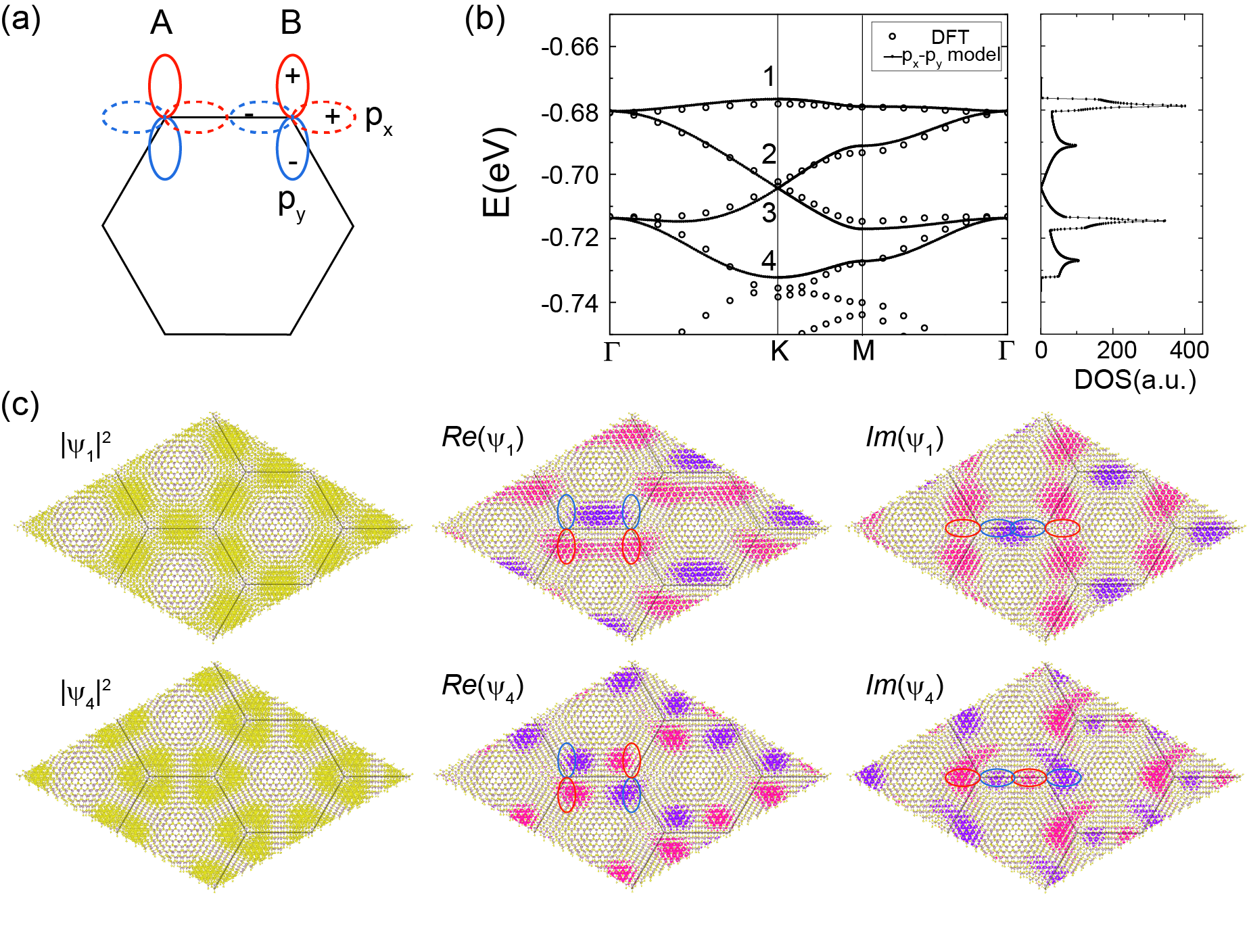}
	  \caption{\captiontitle{p$_{\rm x}$-p$_{\rm y}$ honeycomb model for twisted bilayer MoS$_2$.} (a) Illustration of the model: in a honeycomb lattice composed of sublattices A and B, there are two orthogonal orbitals (p$_{\rm x}$ and p$_{\rm y}$) at each of the two sublattice sites. The solid and the dashed lines denote the p$_{\rm y}$ and the p$_{\rm x}$ orbitals, respectively, and the red and the blue color denotes the positive and the negative side of the orbital, respectively. (b) Fitting the dispersion of the p$_{\rm x}$-p$_{\rm y}$ model to the second set of valence bands of tbMoS$_2$ calculated with DFT for tbMoS$_2$ at $2.65^\circ$. The left panel shows the corresponding density of states displaying the signature four-peak structure. (c) Charge density, real and imaginary parts of the wavefunction calculated with DFT for the states in the two quasi-flat bands 1 and 4 shown in (b). The solutions of the corresponding states from the p$_{\rm x}$-p$_{\rm y}$ model are indicated with the blue and red ovals and agree with  the DFT results.     }
    \label{fig:fig2}
\end{figure}

We first characterize the low-energy electronic properties of twisted bilayer MoS$_2$ using density functional theory (DFT) calculations (see Methods). DFT in particular has established itself as a reliable tool to provide theoretical guidance and to predict the band structure of many twisted bi- and multilayer materials \cite{tran2019evidence,Kerelsky19,Wang19TMD}. However, such a first principles characterization becomes numerically very demanding as the twist angle $\Theta$ approaches small values and the unit cell becomes very large entailing many atoms (of the order of a few thousands and more). Nevertheless, it is that limit in which strong band-narrowing effects and as a consequence prominent effects of correlations are expected. The results of such a characterization are summarized in Fig.~\ref{fig:fig1}. Panel (a) shows the relaxed atomic structure of two sheets of MoS$_2$ in real space, twisted with respect to each other. A Moir\'e interference pattern forms at small twist angle yielding a large unit cell, within which we identify different local patterns of stacking of the two sheets of MoS$_2$, indicated via areas framed by cyan, magenta or purple dashed lines. The local stacking arrangements of the respective areas are given in the right sub-panels of panel (a). In panel (b) we show the  \textit{ab initio} band structure of the twisted material after relaxation, where we find two families of bands which will become increasingly flat and start to detach from all other bands, as the twist angle is lowered. We mark these bands by blue and red color in panel (b), which shows results for decreasing angles from $\Theta=3.16^\circ$ to $2.28^\circ$. The bandwidth of these two energetically-separated groups of bands is summarized in panel (c) of Fig.~\ref{fig:fig1}. We find that the bandwidth of these two bands shrinks drastically as the angle is decreased, yielding band widths of the order of $10$ meV as the angle approaches $\Theta \approx 2^\circ$. Note that these flat bands near the top of the valence bands originate from the states around the $\Gamma$ point in the Brillouin zone of the primitive unit cell of untwisted MoS$_2$, with both S p$_{\rm z}$ and Mo d$_{\rm z^2}$ characters. This is different from the case of twisted WSe$_2$, in which the top valence flat bands originate from the states around the K point in the Brillouin zone of the primitive unit cell \cite{Wang19TMD}. Since in other TMDs, such as MoSe$_2$ and WS$_2$, the top of the valence band in the untwisted bilayer is also located at the $\Gamma$ point in the Brillouin zone \cite{kuc2011,zhang2014direct}, the physics we discussed here transfers also to those materials being twisted.     

The upper bands in Fig.~\ref{fig:fig1} (marked in blue) show a Dirac cone at the K point and behave very similar to the bands found for monolayer  graphene (with the exception of a reduced band width). They are spin degenerate in nature, but feature no additional degeneracy except at certain high symmetry points. Instead, the next set of bands (marked in red) is essential to our work. They too feature a Dirac cone at the K point, but also feature two additional ultra-flat bands at the top and bottom in addition to a band structure similar to graphene. The ratio between the width of the ultra-flat and the flat bands decreases as the angle is decreased, but saturates in our calculations as a twist-angle of $\Theta\approx2.28^\circ$ is approached. We attribute this saturation to lattice relaxation effects; note however that the overall band width keeps decreasing. 

Remarkably, this second family of flat bands is well-described by an effective p$_{\rm x}$-p$_{\rm y}$ tight-binding model on a honeycomb lattice, depicted schematically in Fig.~\ref{fig:fig2} (a), and conveniently described by the following Hamiltonian:
\begin{equation}
\begin{aligned}
  H_0=&\sum\limits_{\left\langle i,j\right\rangle,s} (t_{\sigma} \mathbf{c}^{\dagger}_{i,s}\cdot\mathbf{n}^{\parallel}_{ij} \mathbf{n}^{\parallel}_{ij}\cdot \mathbf{c}_{j,s}-t_{\pi} \mathbf{c}^{\dagger}_{i,s}\cdot\mathbf{n}^{\perp}_{ij} \mathbf{n}^{\perp}_{ij}\cdot \mathbf{c}_{j,s})\,+ \\
 &\sum\limits_{\left\langle \left\langle i,j\right\rangle\right\rangle,s} (t^{N}_{\sigma} \mathbf{c}^{\dagger}_{i,s}\cdot\mathbf{n}^{\parallel}_{ij} \mathbf{n}^{\parallel}_{ij}\cdot \mathbf{c}_{j,s}-t^{N}_{\pi} \mathbf{c}^{\dagger}_{i,s}\cdot\mathbf{n}^{\perp}_{ij} \mathbf{n}^{\perp}_{ij}\cdot \mathbf{c}_{j,s}),
\end{aligned}\label{eq:pxpy}
\end{equation}
where $\mathbf{c}_{i,s}=(c_{i,{\rm x},s},c_{i,{\rm y},s})^{T}$ with $c_{i,{\rm x}({\rm y}),s}$ annihilating an electron with p$_{{\rm x}({\rm y})}$-orbital at site $i$ and with spin $s=\uparrow,\downarrow$. $\left\langle i,j\right\rangle$ ($\left\langle \left\langle i,j\right\rangle\right\rangle$) denotes (next) nearest neighbors. For each sum in Eq.~{\eqref{eq:pxpy}}, the first term describes the $\sigma$ hopping (head to tail) between the p-orbitals and the second term denotes the $\pi$ hopping (shoulder to shoulder). Furthermore,  $\mathbf{n}^{\parallel}_{ij}=(\mathbf{r}_i-\mathbf{r}_j)/|\mathbf{r}_i-\mathbf{r}_j|$, with $\mathbf{r}_i$ being the position of site $i$ and $\mathbf{n}^{\perp}_{ij}=U\mathbf{n}^{\parallel}_{ij}$ with $U$ being the two-dimensional 90 degree rotation matrix $U=\begin{pmatrix}0&-1\\1&0\end{pmatrix}$.  Finally, $t_{\sigma}$ and $t_{\pi}$ ( $t^N_{\sigma}$ and $t^N_{\pi}$) are the nearest neighbor (next nearest neighbor) hopping amplitudes for the $\sigma$-bonding term and $\pi$-bonding term, respectively. Fig.~\ref{fig:fig2}(b) and (c) depict the corresponding dispersions, density of states, and wave functions in comparison to model predictions, illustrating that the four Moir\'e bands at low energies are well-captured by Eq. (\ref{eq:pxpy}) upon the choice of hopping parameters $t_{\pi} = 0.25 t_{\sigma}$, $t^N_{\sigma}=0.07t_{\sigma}$ and $t^N_{\pi}=-0.04t_{\sigma}$. The density of states exhibits a characteristic four van Hove singularities structure, with two originating from the Dirac bands and two stemming from the additional two ultra-flat bands. The small ratio between the nearest neighbor hopping amplitudes $t_{\pi}/t_{\sigma}$ determines the residual small dispersion in the ultra-flat bands we report. This ratio is controllable by the twist angle, which is summarized in the inset of Fig.~\ref{fig:fig1} (c). All these parameters are related to the interlayer Moir\'e potential and thus expected to be also affected and controllable by the uniaxial pressure perpendicular to the layers as demonstrated for twisted bilayer graphene \cite{Yankowitz18}.

The flat band wavefunctions consist of atomic wavefunctions from the p$_{\rm z}$ orbital on S~atoms and the d$_{\rm z^2}$ orbital on Mo atoms. Modulated by the Moir\'e potential, the weighting of the atomic wavefunctions and their modulus square (i.e., charge density) vary at different atomic sites across the whole supercell, showing distinct patterns for different flat band states at the K point in the supercell Brillouin zone as shown in Panel (c) of Fig.~\ref{fig:fig2}. These patterns of the charge density as well as the real and the imaginary part of the total wavefunctions obtained from DFT show features consistent with those of the p$_{\rm x}$-p$_{\rm y}$ Hamiltonian of Eq.~\eqref{eq:pxpy}. Interestingly, the charge density distribution of the top ultra-flat band state displays a Kagome lattice structure. We have thus unambiguously established twisted MoS$_2$ to be a candidate system to realize a p$_{\rm x}$-p$_{\rm y}$ model on the honeycomb lattice with strongly asymmetric hoppings $t_{\sigma}$ and $t_\pi$, giving rise to a new set of ultra-flat bands.

\begin{figure}
    \centering
    \includegraphics[width=\textwidth]{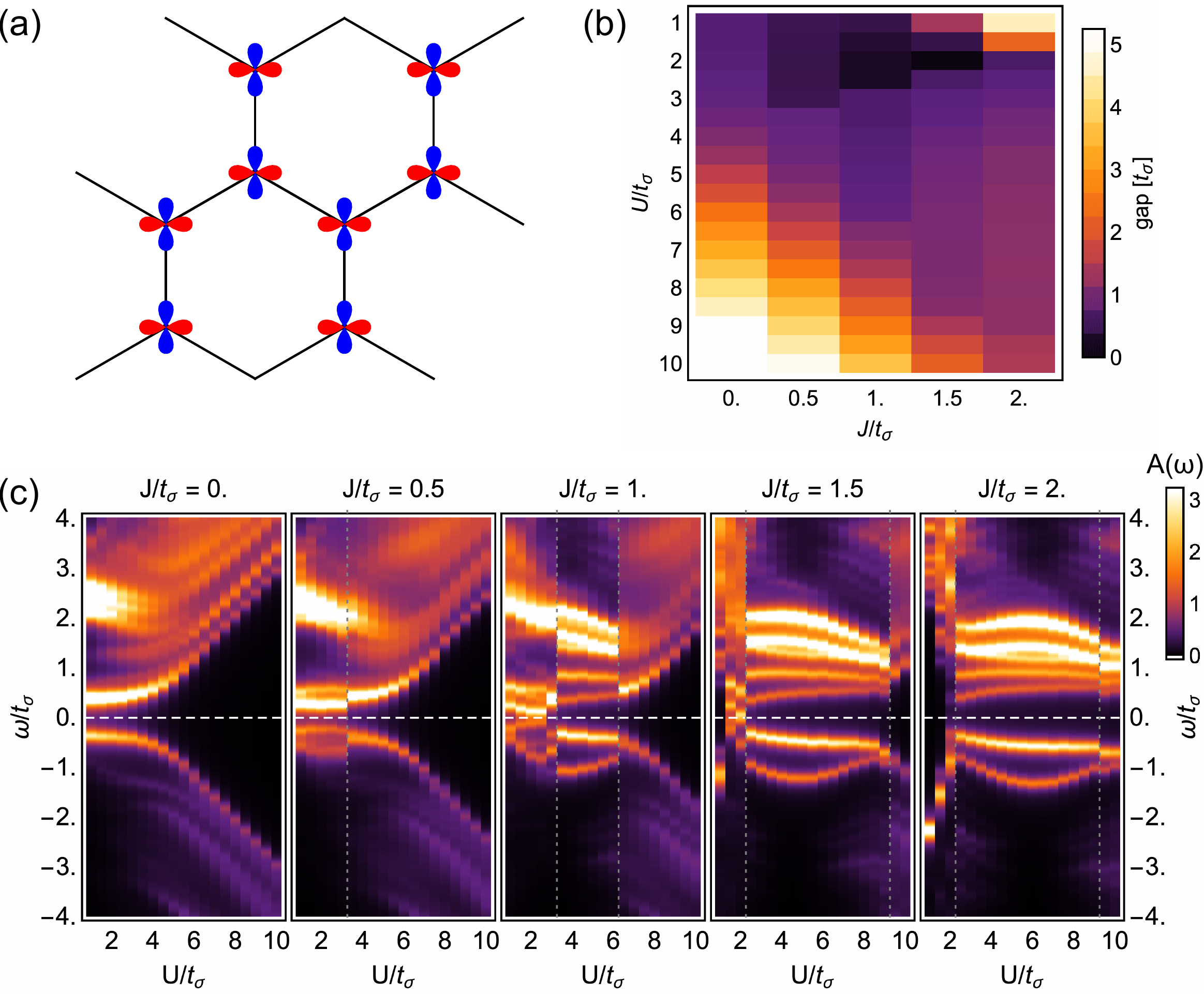}
    \caption{\textbf{Charge gap and correlations for twisted bilayer MoS$_2$ at vanishing temperature.} (a) depicts the 16-orbital cluster geometry employed for exact diagonalization of the Hubbard-Kanamori Hamiltonian. (b) depicts the charge gap as a function of Hubbard $U$ and Hund's exchange $J$ interactions, calculated for the 16-orbital cluster and extracted from (c) the local density of states, which is readily accessible via scanning tunnelling microscopy. A well-defined charge gap develops beyond $U/t_\sigma \sim 4$ at small $J$ that scales linearly with the Hubbard interaction $U$. Vertical gray dotted lines indicate phase transitions to charge-ordered states at large $J/U$, coinciding with a closing of the charge gap.}
    \label{fig:ED2}
\end{figure}


\begin{figure}
    \centering
	\includegraphics[width = \columnwidth]{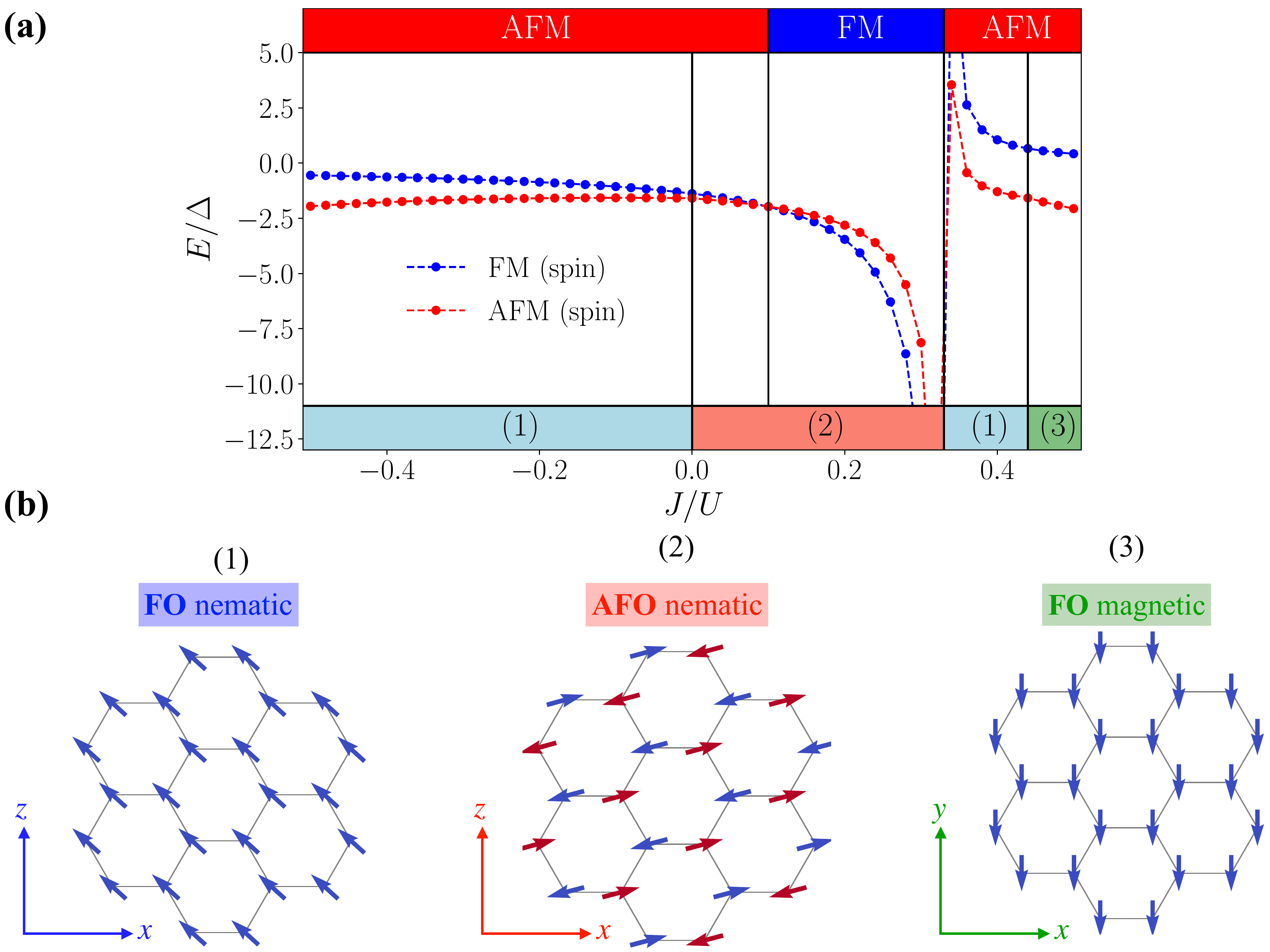}
    \caption{\captiontitle{Magnetic phase diagram for twisted bilayer MoS$_2$.} (a) Classical ground state energy per orbital in units of $\Delta = t_{\sigma}^{2} / U$, assuming ferro- (blue) or antiferromagnetic (red) order for the spin degrees of freedom. We take the \textit{ab initio} parameters, $t_{\pi} = 0.25 t_{\sigma}$ and use an iterative energy minimization. The lower panel determines the phase boundaries for the orbital degrees of freedom given the energetically more favorable spin order shown in the top panel. At $J/U = 0.1$ we find the spin order to change from AFM to FM, with AFO nematic order for the orbital degrees of freedom remaining stable in agreement with \cite{PhysRevB.100.205131}. (b) Configurations of orbital vectors found at the end of iterative minimization. Note that we display the projection of $\boldsymbol{\tau}$ to the plane in $\mathbb{R}^3$ (indicated by the axis shown in the bottom left), such that nematic states with finite contributions only in $xz$ direction ((1) \& (2)) can be distinguished from magnetic states (3) which point perpendicular, i.e along the $y$-axis.}
    \label{fig:phase_diagram}
\end{figure}

\subsection*{Correlations and magnetic properties}
We now study the role of electronic interactions. As the highly-anisotropic p$_{\rm x}$-p$_{\rm y}$ orbital structure constitutes the essential novelty of twisted bilayer MoS$_2$, we focus on quarter filling where orbital fluctuations can be expected to be crucial. This filling fraction is straightforwardly accessible in experiment via back gating, and we defer a discussion of the half-filled case to the Supplementary Material. To proceed, we assume purely local electronic interactions, which can be generically parameterized in terms of the Hubbard-Kanamori Hamiltonian:
\begin{equation}
    H_U = U \sum_{i, \alpha} n_{i \alpha \uparrow} n_{i \alpha \downarrow} 
    + (U - 2J) \sum_{i} n_{i x} n_{i y} 
    + J \sum_{i, s, s'} c^{\dagger}_{i x s} c^{\dagger}_{i y s'} c^{\phantom{\dagger}}_{i x s'}  c^{\phantom{\dagger}}_{i y s} 
    + J \sum_{i, \alpha \neq \beta} c^{\dagger}_{i \alpha \uparrow} c^{\dagger}_{i \alpha \downarrow} c^{\phantom{\dagger}}_{i \beta \downarrow}  c^{\phantom{\dagger}}_{i \beta \uparrow} \label{eq:HubbardKanamori}
\end{equation}
for two orbitals with rotational symmetry. Furthermore, our DFT calculations suggest $t_{\pi} \approx 0.25 t_{\sigma}$ and only weak next-nearest neighbor hopping at small twist angles; we therefore neglect next-nearest-neighbor hopping in the analysis below.  

Fig. \ref{fig:ED2} (c) depicts the local density of states as a function of Hubbard $U$ and Hund's exchange $J$ interactions, calculated via an exact diagonalization study of Eqs. (\ref{eq:pxpy}) and (\ref{eq:HubbardKanamori}) for a cluster depicted schematically in (a). Clear evidence of a charge gap beyond $U/t_\sigma \sim 4$ at small $J$ signifies the onset of a correlated insulator which could be directly observed via transport and scanning tunnelling microscopy. The behavior of the gap is depicted in Fig. \ref{fig:ED2}(b) as a function of $U,J$ and signifies that charge fluctuations are strongly suppressed for large $U$.

In this regime, a natural follow-up questions concerns possible orderings of the orbital and magnetic degrees of freedom. The corresponding strong-coupling Kugel-Khomskii Hamiltonian \cite{tokura2000orbital,corboz2012,khomskii2003orbital} for the p$_{\rm x}$-p$_{\rm y}$ model at quarter filling is given in Ref.~\cite{PhysRevB.90.094422,PhysRevB.98.245103,PhysRevB.100.205131} and reads:
\begin{align}
\label{eq:HSC}
H = \sum_{\langle ij \rangle} & \frac{1}{U - 3J} \xi^{1}_{ij} \left[t_{\sigma} t_{\pi} \bar{Q}_{ij} - (t_{\sigma}^{2} + t_{\pi}^{2}) (P_{ij}^{xy} + P_{ij}^{yx}) \right] \notag \\
- & \frac{1}{U + J} \xi_{ij}^{0} \left[ t_{\sigma} t_{\pi} Q_{ij} + 2 t_{\sigma}^{2} P_{ij}^{xx} + 2 t_{\pi}^{2} P_{ij}^{yy} \right] \notag \\
+ & \frac{1}{U - J} \xi_{ij}^{0} \left[ t_{\sigma} t_{\pi} (Q_{ij} - \bar{Q}_{ij}) - 2 t_{\sigma}^{2} P_{ij}^{xx} - 2 t_{\pi}^{2} P_{ij}^{yy} - (t_{\sigma}^{2} + t_{\pi}^{2}) (P_{ij}^{xy} + P_{ij}^{yx}) \right] \, . 
\end{align}
Here, $\xi_{ij}^{1} = 3/4 + \mathbf{S}_i \mathbf{S}_j$ denotes the projector onto triplet states, whereas $\xi_{ij}^{0} = 1/4 - \mathbf{S}_i \mathbf{S}_j$ selects the singlet spin states instead. Note that the orbital operators, for example $Q_{ij}$, are bond dependent, giving rise to a strong spatial anisotropy of the resulting spin-orbit model. Details about their definitions can be found in the Methods section.

To study its ground state phase diagram using the \textit{ab initio} parameters found in the previous section, we employ a mean-field analysis of competing orbital orderings with ferromagnetic and antiferromagnetic spin order. To this end, we note that on the bipartite honeycomb lattice the SU($2$) invariant spin sector would, on its own, order either ferro- or antiferromagnetically, depending on the sign of the exchange couplings. As an Ansatz we therefore assume that one of the respective states is stabilized and decouple the spin from the orbital degrees of freedom by replacing $\mathbf{S}_{i} \mathbf{S}_{j}$ with its expectation value $\langle \mathbf{S}_{i} \mathbf{S}_{j} \rangle = \pm 1/4$ such that $\xi_{ij}^{1} = 1, \xi_{ij}^{0} = 0$ for ferromagnetic spin order and $\xi_{ij}^{1} = \xi_{ij}^{0} = 1/2$ for Ne\'el order.

After such a mean-field decoupling corresponding to the ground state in the spin sector, we analyze the ground states of the resulting Hamiltonian for the orbital degrees of freedom, which we approximate as classical vectors. We use an iterative energy minimization combined with simulated annealing techniques (see Methods) to converge the mean field equations and find the phase diagram summarized in Fig.~\ref{fig:phase_diagram}.
 Panel (a) shows the energy of ferromagnetic and antiferromagnetic spin configurations from which the magnetic phase diagram can be read off. This is given in the upper part of the plot and we find antiferromagnetic ordering with an intermittent ferromagnetic phase at intermediate ratios of $0.1 < J/U < 1/3$.  In the lower part of the plot we show the corresponding subsidiary orbital order. From our simulations we identify three different configurations of orbital vectors $\boldsymbol{\tau}$, which can be classified according to their projection on a single definite plane in  space, shown in the lower left of the plots: (1) ferro-orbital (FO) nematic order, where the vectors on all lattice sites align in parallel to the xz-plane. Quantum mechanically, finite values of $\langle \tau^{x/z}_{i} \rangle$ indicate an imbalance of the occupation of p$_{\rm x}$ and p$_{\rm y}$ orbitals, breaking rotation symmetry and thereby motivating the notion of a nematic state. (2) AFO nematic order; each vector is aligned anti-parallel with its nearest neighbors corresponding to $\langle \tau^{x/z}_{i} \rangle \neq 0$ on each sublattice, but without finite projections $\tau_{i}^{y}$ on individual sites. (3) FO magnetic order; all vectors order along the $y$-axis, such that $\langle \tau^{y}_{i} \rangle \neq 0$, which, in the quantum mechanical system, would indicate time-reversal symmetry breaking. 
 The inclusion of quantum fluctuations can change this picture and more exotic ground states may emerge. For example, for our \textit{ab initio} band structure parameters, a noncollinear spin dimer phase is predicted in a certain range of interaction couplings and even a quantum spin-orbital liquid is found in its proximity~\cite{PhysRevB.100.205131}.

\section*{Discussion}
We have established that twisted bilayer MoS$_2$ is a promising platform to realize the orbital anisotropic p$_{\rm x}$-p$_{\rm y}$ Hubbard model by employing large scale {\it ab initio}
calculations. We find that families of flat bands emerge where the first family of flat bands shows s-orbital character and the second family is an intriguing realization of a strongly asymmetric p$_{\rm x}$-p$_{\rm y}$ Hubbard model both on a honeycomb lattice, adding a lattice with non-trivial almost perfectly-flat bands due to destructive interference to the growing list of systems that can be engineered in twisted heterostructures. At even smaller angle the sequence in the family of flat bands found with respect to their orbital character continues and a preliminary study shows 
that the next family would exhibit d-orbital character on the honeycomb  lattice. Such a lattice would effectively realize a multi-orbital generalization of a Kagome lattice -- a prototypical model for quantum spin liquids. However, at such small angles strong relaxation is likely to become dominant, prohibiting access to this regime and  potentially spoiling its experimental realization. Currently the {\it ab initio} characterization of such small angles is numerically too exhaustive and this work sparks a direct need for novel computational methods to tackle this question.

Furthermore, our combined exact diagonalization and strong-coupling expansion approaches classify the magnetic and orbital phase diagrams, however, inclusion of quantum fluctuations stipulates an intriguing avenue of future theoretical research. 
In addition, by proximity or variations in the chemical composition of the twisted bilayer, it might be possible to induce spin-orbit coupling splitting of the ultra-flat bands at the top and bottom of the asymmetric p$_{\rm x}$-p$_{\rm y}$ dispersion. Such a band gap opening would induce interesting topological properties \cite{PhysRevB.90.075114} in a highly tunable materials setting.

\clearpage

\appendix

\section*{Methods} 

{\it Details on \textit{ab initio} calculations}

We calculate the electronic properties of twisted bilayer MoS$_2$ with \textit{ab initio} methods based on density functional theory (DFT) as implemented in the Vienna {\it ab initio} Simulation Package (VASP) \cite{kresse93ab}. We employ plane wave basis sets with an energy cutoff of 550 eV and pseudopotentials as constructed with the projector augmented wave (PAW) method \cite{blochl94}. The exchange-correlation functionals are treated at the generalized gradient approximations (GGA) level \cite{pbe}. The supercell lattice constants are chosen such that they correspond to 3.161 {\AA} for the 1x1 primitive cell of MoS$_2$. Vacuum spacing larger than 15 {\AA} is introduced to avoid artificial interaction between the periodic images along the z direction. Because of the large supercells, a 1x1x1 k-grid is employed for the ground state and the relaxation calculations. For all the calculations, all the atoms are relaxed until the force on each atom is less than 0.01 eV/{\AA}. Van der Waals corrections are considered with the method of Tkatchenko and Scheffler \cite{tsmethod09}. \\

\noindent
{\it Details on exact diagonalization}

Exact diagonalization calculations were performed for the electronic tight-binding model in Eq. (\ref{eq:pxpy}) with Hubbard Kanamori interactions defined in Eq. (\ref{eq:HubbardKanamori}). All calculations were performed for the total momentum $\mathbf{K}_{\rm tot} = 0$ and total spin $S_z = 0$ sector, for a two-orbital eight-site cluster with periodic boundary conditions. Rotationally-symmetric Kanamori interactions are adopted, with $U' = U - 2J$. As the magnitudes of the Hubbard $U$ and Hund's exchange $J$ interactions cannot be reliably predicted for a Moi\'e super cell from first principles, all presented results are shown as a function of $U$, $J$. Calculations of the single-particle Green's functions and local density of states are performed using the Lanczos method and continued-fraction representation, and a spectral broadening (imaginary part of the self energy) of $\eta = 0.1$ is imposed.\\

\noindent
{\it Details on strong-coupling Hamiltonian}
\label{sc_details}

Here we give the definitions of the operators used in \eqref{eq:HSC} and discussed in \cite{PhysRevB.98.245103}. The operators $Q_{ij}$ and $\bar{Q}_{ij}$ describe processes where orbital occupations of sites $i$ and $j$ are reversed, that is they are defined as
\begin{align}
    Q_{ij}       &= \left(\tau_{i}^{+} \tau_{j}^{+} + \tau_{i}^{-} \tau_{j}^{-}\right) / 2 \notag \\
    \bar{Q}_{ij} &= \left(\tau_{i}^{+} \tau_{j}^{-} + \tau_{i}^{-} \tau_{j}^{+}\right) / 2 \,,
\end{align}
with 
\begin{align}
    \tau_{i}^{\pm} = \mathbf{n}^{\perp}_{ij} \boldsymbol{\tau}_i \pm i \tau_{i}^{y} \,,
\end{align}
where $\boldsymbol{\tau}_i = (\tau_{i}^{z}, \tau_{i}^{x}, \tau_{i}^{y})^{T}$. The orbital projection operators can then be expressed as
\begin{align}
    P_{ij}^{xx} &= (1 + \mathbf{n}^{\parallel}_{ij} \boldsymbol{\tau}_{i}) (1 + \mathbf{n}^{\parallel}_{ij} \boldsymbol{\tau}_{j}) / 4 \notag \\
    P_{ij}^{yy} &= (1 - \mathbf{n}^{\parallel}_{ij} \boldsymbol{\tau}_{i}) (1 - \mathbf{n}^{\parallel}_{ij} \boldsymbol{\tau}_{j}) / 4 \notag \\
    P_{ij}^{xy} &= (1 + \mathbf{n}^{\parallel}_{ij} \boldsymbol{\tau}_{i}) (1 - \mathbf{n}^{\parallel}_{ij} \boldsymbol{\tau}_{j}) / 4 \notag \\
    P_{ij}^{yx} &= (1 - \mathbf{n}^{\parallel}_{ij} \boldsymbol{\tau}_{i}) (1 + \mathbf{n}^{\parallel}_{ij} \boldsymbol{\tau}_{j}) / 4 \,.
\end{align}
where e.g. $P_{ij}^{xx}$ selects states where the superposition $(p_x \mathbf{e}_x + p_y \mathbf{e}_y) \mathbf{n}^{\parallel}_{ij}$ is occupied on nearest-neighbor sites $i$ and $j$ connected by the bond $\mathbf{n}^{\parallel}_{ij}$.\\

\noindent
{\it Details on minimization procedure for classical Hamiltonian}
 
\begin{figure}
    \centering
	\includegraphics[width = \columnwidth]{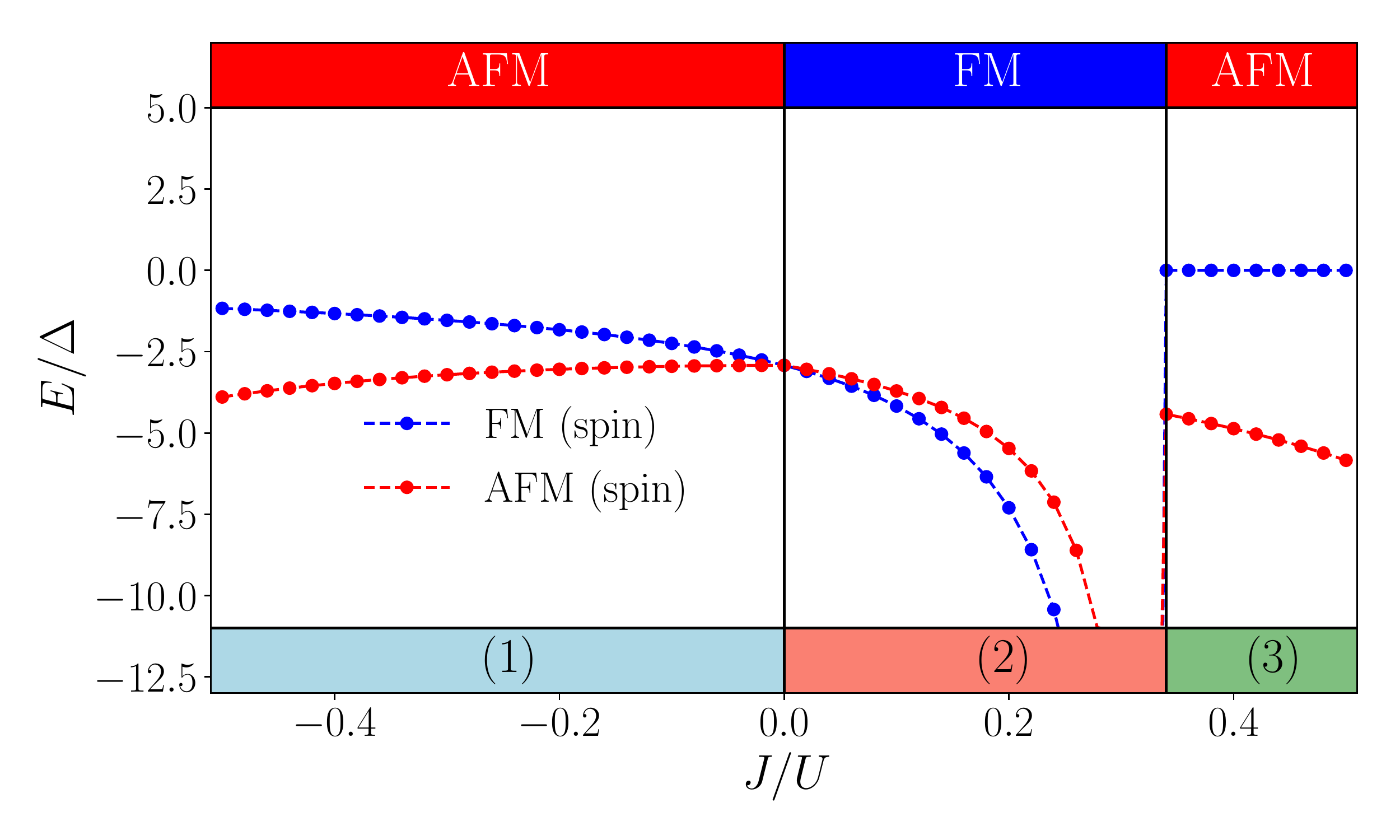}
    \caption{\textbf{Magnetic phase diagram of the strong-coupling Hamiltonian \eqref{eq:HSC} in the isotropic limit} $t_{\sigma} = t_{\pi}$. Our results from iterative minimization are in agreement with \cite{PhysRevB.98.245103}, stabilizing FO nematic order (1) for $J / U < 0$ (see main text) and FO magnetic order (3) (see main text) for $J / U > 1 / 3$. In the intermediate range of parameters $0 < J / U < 1/3$ the ferromagnetic spin sector is selected, such that, due to vanishing $\xi_{ij}^{0}$, rotation invariance is restored for the orbital vectors, giving rise to AFM order (2) but without any preferred axis in euclidean space.}
    \label{fig:benchmark}
\end{figure}

Metropolis Monte Carlo simulations are a prime tool for the investigation of classical spin models, since they allow for off-diagonal, spatially anisotropic spin couplings to be included, even when one-spin terms, such as magnetic fields, are involved. Here we employ a special variant of the algorithm to the mean-field version of \eqref{eq:HSC}, keeping in mind that the `spins' used in the simulation are approximations to orbital operators $\boldsymbol{\tau}$. \\
First, a lattice site $i$ is randomly chosen, and its respective gradient field $\mathbf{h}_{i} = \nabla_{i} H$ is computed for the current spin configuration $\{ \boldsymbol{\tau}_{i} \}$. Second, a random orientation $\boldsymbol{\tau}'_{i}$ for the vector at site $i$ is proposed and the weight
\begin{align}
    g = \mathrm{min} \left( e^{-\beta (\boldsymbol{\tau}'_{i} - \boldsymbol{\tau}_{i}) \mathbf{h}_{i}}, 1 \right) \,,
\end{align}
is computed for an effective inverse temperature $\beta$. Performing several Metropolis updates with increasing values of $\beta$ we are able to efficiently lower the energy of a random initial configuration, minimizing the odds to converge to a local minimum by only allowing optimal updates (i.e. $\boldsymbol{\tau}_{i} = -\mathbf{h}_{i}$) right from the start. After $N_a$ sweeps over the full lattice, the so-obtained configuration is ameliorated by $N_o$ optimization sweeps, where the randomly selected spin is rotated anti-parallel to the local gradient field such that the energy is deterministically lowered in every step and we converge as close to the global energy minimum as possible. Hence, this algorithm is reminiscent of Monte Carlo simulations with simulated annealing, but at zero temperature where thermal fluctuations are frozen out. \\
To benchmark our implementation we have carried out the minimization procedure in the isotropic limit $t_{\sigma} = t_{\pi}$ for $N_a = N_o = 10^5$, where the optimization sweeps are terminated when the energy change after one sweep, $\epsilon$, becomes small (usually $\epsilon \leq 10^{-10}$). Mapping out the phase diagram for both the FM, $\langle \mathbf{S}_i \mathbf{S}_j \rangle = 1/4$, as well as the AFM, $\langle \mathbf{S}_i \mathbf{S}_j \rangle = -1/4$, spin sector on a lattice with $N = 1250$ spins subject to periodic boundary conditions we find the result in Fig. \ref{fig:benchmark}, which is consistent with the one presented in \cite{PhysRevB.98.245103}. For $J < 0$ the AFM spin sector has lower energy, with the orbitals forming a ferro-orbital (FO) nematic state where $\langle \tau^{x/z}_{i} \rangle \neq 0$ and $\langle \tau^{y}_{i} \rangle = 0$. For $J > 0$ one finds the FM spin sector (for which the orbital degrees of freedom restore their rotation invariance) to dominate as long as $J < 1/3$, where the AFM sector takes over again and establishes a FO magnetic state, i.e. $\langle \tau^{x/z}_{i} \rangle = 0$ and $\langle \tau^{y}_{i} \rangle \neq 0$.

\bibliographystyle{ScienceAdvances}
\bibliography{moirebib}

\begin{thebibliography}{10}

\bibitem{liu2020disassembling}
F.~Liu, W.~Wu, Y.~Bai, S.~H. Chae, Q.~Li, J.~Wang, J.~Hone, X.-Y. Zhu,
  Disassembling 2d van der waals crystals into macroscopic monolayers and
  reassembling into artificial lattices.
\newblock {\it Science\/} {\bf 367}, 903--906 (2020).

\bibitem{cao2018a}
Y.~Cao, V.~Fatemi, S.~Fang, K.~Watanabe, T.~Taniguchi, E.~Kaxiras,
  P.~Jarillo-Herrero, Unconventional superconductivity in magic-angle graphene
  superlattices.
\newblock {\it Nature\/} {\bf 556}, 43–50 (2018).

\bibitem{cao2018b}
Y.~Cao, V.~Fatemi, A.~Demir, S.~Fang, S.~L. Tomarken, J.~Y. Luo, J.~D.
  Sanchez-Yamagishi, K.~Watanabe, T.~Taniguchi, E.~Kaxiras, R.~C. Ashoori,
  P.~Jarillo-Herrero, Correlated insulator behaviour at half-filling in
  magic-angle graphene superlattices.
\newblock {\it Nature\/} {\bf 556}, 80–84 (2018).

\bibitem{Yankowitz18}
M.~Yankowitz, S.~Chen, H.~Polshyn, Y.~Zhang, K.~Watanabe, T.~Taniguchi,
  D.~Graf, A.~F. Young, C.~R. Dean, Tuning superconductivity in twisted bilayer
  graphene.
\newblock {\it Science\/} {\bf 363}, 1059--1064 (2019).

\bibitem{Kerelsky18}
A.~Kerelsky, L.~J. McGilly, D.~M. Kennes, L.~Xian, M.~Yankowitz, S.~Chen,
  K.~Watanabe, T.~Taniguchi, J.~Hone, C.~Dean, A.~Rubio, A.~N. Pasupathy,
  Maximized electron interactions at the magic angle in twisted bilayer
  graphene.
\newblock {\it Nature\/} {\bf 572}, 95-100 (2019).

\bibitem{Choi19}
Y.~Choi, J.~Kemmer, Y.~Peng, A.~Thomson, H.~Arora, R.~Polski, Y.~Zhang, H.~Ren,
  J.~Alicea, G.~Refael, F.~von Oppen, K.~Watanabe, T.~Taniguchi, S.~Nadj-Perge,
  Electronic correlations in twisted bilayer graphene near the magic angle.
\newblock {\it Nat. Phys.\/} {\bf 15}, 1174-1180 (2019).

\bibitem{Jin18}
C.~Jin, E.~C. Regan, A.~Yan, M.~Iqbal Bakti~Utama, D.~Wang, S.~Zhao, Y.~Qin,
  S.~Yang, Z.~Zheng, S.~Shi, K.~Watanabe, T.~Taniguchi, S.~Tongay, A.~Zettl,
  F.~Wang, Observation of moir{\'e} excitons in wse2/ws2 heterostructure
  superlattices.
\newblock {\it Nature\/} {\bf 567}, 76-80 (2019).

\bibitem{tran2019evidence}
K.~Tran, G.~Moody, F.~Wu, X.~Lu, J.~Choi, K.~Kim, A.~Rai, D.~A. Sanchez,
  J.~Quan, A.~Singh, {\it et~al.\/}, Evidence for moir{\'e} excitons in van der
  waals heterostructures.
\newblock {\it Nature\/} {\bf 567}, 71--75 (2019).

\bibitem{alexeev2019resonantly}
E.~M. Alexeev, D.~A. Ruiz-Tijerina, M.~Danovich, M.~J. Hamer, D.~J. Terry,
  P.~K. Nayak, S.~Ahn, S.~Pak, J.~Lee, J.~I. Sohn, {\it et~al.\/}, Resonantly
  hybridized excitons in moir{\'e} superlattices in van der waals
  heterostructures.
\newblock {\it Nature\/} {\bf 567}, 81--86 (2019).

\bibitem{seyler2019signatures}
K.~L. Seyler, P.~Rivera, H.~Yu, N.~P. Wilson, E.~L. Ray, D.~G. Mandrus, J.~Yan,
  W.~Yao, X.~Xu, Signatures of moir{\'e}-trapped valley excitons in mose 2/wse
  2 heterobilayers.
\newblock {\it Nature\/} {\bf 567}, 66--70 (2019).

\bibitem{Xian18}
L.~Xian, D.~M. Kennes, N.~Tancogne-Dejean, M.~Altarelli, A.~Rubio, Multiflat
  bands and strong correlations in twisted bilayer boron nitride:
  Doping-induced correlated insulator and superconductor.
\newblock {\it Nano Lett.\/} {\bf 19}, 4934-4940 (2019).

\bibitem{Naik18}
M.~H. Naik, M.~Jain, Ultraflatbands and shear solitons in moir\'e patterns of
  twisted bilayer transition metal dichalcogenides.
\newblock {\it Phys. Rev. Lett.\/} {\bf 121}, 266401 (2018).

\bibitem{Wang19TMD}
L.~Wang, E.-M. Shih, A.~Ghiotto, L.~Xian, D.~A. Rhodes, C.~Tan, M.~Claassen,
  D.~M. Kennes, Y.~Bai, B.~Kim, K.~Watanabe, T.~Taniguchi, X.~Zhu, J.~Hone,
  A.~Rubio, A.~Pasupathy, C.~R. Dean, Magic continuum in twisted bilayer wse2.
\newblock {\it Preprint at https://arxiv.org/abs/1910.12147\/}  (2019).

\bibitem{Scuri19}
G.~Scuri, T.~I. Andersen, Y.~Zhou, D.~S. Wild, J.~Sung, R.~J. Gelly,
  D.~Bérubé, H.~Heo, L.~Shao, A.~Y. Joe, A.~M.~M. Valdivia, T.~Taniguchi,
  K.~Watanabe, M.~Lončar, P.~Kim, M.~D. Lukin, H.~Park, Electrically tunable
  valley dynamics in twisted wse$_2$/wse$_2$ bilayers.
\newblock {\it Preprint at https://arxiv.org/abs/1910.12147\/}  (2019).

\bibitem{Kerelsky19}
A.~Kerelsky, C.~Rubio-Verdú, L.~Xian, D.~M. Kennes, D.~Halbertal, N.~Finney,
  L.~Song, S.~Turkel, L.~Wang, K.~Watanabe, T.~Taniguchi, J.~Hone, C.~Dean,
  D.~Basov, A.~Rubio, A.~N. Pasupathy, Moiré-less correlations in abca
  graphene.
\newblock {\it Preprint at https://arxiv.org/abs/1911.00007\/}  (2019).

\bibitem{Andersen19}
T.~I. Andersen, G.~Scuri, A.~Sushko, K.~D. Greve, J.~Sung, Y.~Zhou, D.~S. Wild,
  R.~J. Gelly, H.~Heo, K.~Watanabe, T.~Taniguchi, P.~Kim, H.~Park, M.~D. Lukin,
  Moiré excitons correlated with superlattice structure in twisted
  wse$_2$/wse$_2$ homobilayers.
\newblock {\it Preprint at https://arxiv.org/abs/1912.06955\/}  (2019).

\bibitem{Kennes2020}
D.~M. Kennes, L.~Xian, M.~Claassen, A.~Rubio, One-dimensional flat bands in
  twisted bilayer germanium selenide.
\newblock {\it Nature Communications\/} {\bf 11}, 1124 (2020).

\bibitem{lu2019superconductors}
X.~Lu, P.~Stepanov, W.~Yang, M.~Xie, M.~A. Aamir, I.~Das, C.~Urgell,
  K.~Watanabe, T.~Taniguchi, G.~Zhang, {\it et~al.\/}, {Superconductors,
  Orbital Magnets, and Correlated States in Magic Angle Bilayer Graphene} .

\bibitem{stepanov2019interplay}
P.~Stepanov, I.~Das, X.~Lu, A.~Fahimniya, K.~Watanabe, T.~Taniguchi, F.~H.~L.
  Koppens, J.~Lischner, L.~Levitov, D.~K. Efetov, The interplay of insulating
  and superconducting orders in magic-angle graphene bilayers (2019).

\bibitem{arora2020superconductivity}
H.~S. Arora, R.~Polski, Y.~Zhang, A.~Thomson, Y.~Choi, H.~Kim, Z.~Lin, I.~Z.
  Wilson, X.~Xu, J.-H. Chu, K.~Watanabe, T.~Taniguchi, J.~Alicea,
  S.~Nadj-Perge, Superconductivity without insulating states in twisted bilayer
  graphene stabilized by monolayer wse$_2$ (2020).

\bibitem{chen2020nature}
G.~Chen, A.~L. Sharpe, E.~J. Fox, Y.-H. Zhang, S.~Wang, L.~Jiang, B.~Lyu,
  H.~Li, K.~Watanabe, T.~Taniguchi, {\it et~al.\/}, Tunable correlated chern
  insulator and ferromagnetism in a moir{\'e} superlattice.
\newblock {\it Nature\/} {\bf 579}, 56--61 (2020).

\bibitem{liu2019spin}
X.~Liu, Z.~Hao, E.~Khalaf, J.~Y. Lee, K.~Watanabe, T.~Taniguchi, A.~Vishwanath,
  P.~Kim, {Spin-polarized Correlated Insulator and Superconductor in Twisted
  Double Bilayer Graphene} .

\bibitem{shen2019observation}
C.~Shen, N.~Li, S.~Wang, Y.~Zhao, J.~Tang, J.~Liu, J.~Tian, Y.~Chu,
  K.~Watanabe, T.~Taniguchi, {\it et~al.\/}, {Observation of superconductivity
  with Tc onset at 12K in electrically tunable twisted double bilayer graphene}
  .

\bibitem{cao2019electric}
Y.~Cao, D.~Rodan-Legrain, O.~Rubies-Bigord{\`a}, J.~M. Park, K.~Watanabe,
  T.~Taniguchi, P.~Jarillo-Herrero, {Electric Field Tunable Correlated States
  and Magnetic Phase Transitions in Twisted Bilayer-Bilayer Graphene} .

\bibitem{he2020}
M.~He, Y.~Li, J.~Cai, Y.~Liu, K.~Watanabe, T.~Taniguchi, X.~Xu, M.~Yankowitz,
  Tunable correlation-driven symmetry breaking in twisted double bilayer
  graphene (2020).

\bibitem{tutuc2019}
G.~W. Burg, J.~Zhu, T.~Taniguchi, K.~Watanabe, A.~H. MacDonald, E.~Tutuc,
  Correlated insulating states in twisted double bilayer graphene.
\newblock {\it Phys. Rev. Lett.\/} {\bf 123}, 197702 (2019).

\bibitem{liu_2014}
Z.~Liu, F.~Liu, Y.-S. Wu, Exotic electronic states in the world of flat bands:
  From theory to material.
\newblock {\it Chinese Physics B\/} {\bf 23}, 077308 (2014).

\bibitem{leykam2018artificial}
D.~Leykam, A.~Andreanov, S.~Flach, Artificial flat band systems: from lattice
  models to experiments.
\newblock {\it Advances in Physics: X\/} {\bf 3}, 1473052 (2018).

\bibitem{mak2010}
K.~F. Mak, C.~Lee, J.~Hone, J.~Shan, T.~F. Heinz, Atomically thin
  ${\mathrm{mos}}_{2}$: A new direct-gap semiconductor.
\newblock {\it Phys. Rev. Lett.\/} {\bf 105}, 136805 (2010).

\bibitem{wang2018synthesis}
H.~Wang, C.~Li, P.~Fang, Z.~Zhang, J.~Z. Zhang, Synthesis, properties, and
  optoelectronic applications of two-dimensional mos 2 and mos 2-based
  heterostructures.
\newblock {\it Chemical Society Reviews\/} {\bf 47}, 6101--6127 (2018).

\bibitem{naik2019origin}
M.~H. Naik, S.~Kundu, I.~Maity, M.~Jain, Origin and evolution of ultraflatbands
  in twisted bilayer transition metal dichalcogenides: Realization of
  triangular quantum dots (2019).

\bibitem{wu2007prl}
C.~Wu, D.~Bergman, L.~Balents, S.~Das~Sarma, Flat bands and wigner
  crystallization in the honeycomb optical lattice.
\newblock {\it Phys. Rev. Lett.\/} {\bf 99}, 070401 (2007).

\bibitem{wu2008prb}
C.~Wu, S.~Das~Sarma, ${p}_{x,y}$-orbital counterpart of graphene: Cold atoms in
  the honeycomb optical lattice.
\newblock {\it Phys. Rev. B\/} {\bf 77}, 235107 (2008).

\bibitem{bistritzer2011}
R.~Bistritzer, A.~H. MacDonald, Moir{\'e} bands in twisted double-layer
  graphene.
\newblock {\it Proceedings of the National Academy of Sciences\/} {\bf 108},
  12233--12237 (2011).

\bibitem{PhysRevB.98.045103}
N.~F.~Q. Yuan, L.~Fu, Model for the metal-insulator transition in graphene
  superlattices and beyond.
\newblock {\it Phys. Rev. B\/} {\bf 98}, 045103 (2018).

\bibitem{PhysRevB.99.195455}
H.~C. Po, L.~Zou, T.~Senthil, A.~Vishwanath, Faithful tight-binding models and
  fragile topology of magic-angle bilayer graphene.
\newblock {\it Phys. Rev. B\/} {\bf 99}, 195455 (2019).

\bibitem{PhysRevResearch.1.033072}
S.~Carr, S.~Fang, H.~C. Po, A.~Vishwanath, E.~Kaxiras, Derivation of wannier
  orbitals and minimal-basis tight-binding hamiltonians for twisted bilayer
  graphene: First-principles approach.
\newblock {\it Phys. Rev. Research\/} {\bf 1}, 033072 (2019).

\bibitem{TBGFRG}
D.~M. Kennes, J.~Lischner, C.~Karrasch, Strong correlations and $d+\mathit{id}$
  superconductivity in twisted bilayer graphene.
\newblock {\it Phys. Rev. B\/} {\bf 98}, 241407 (2018).

\bibitem{zhang2014}
G.-F. Zhang, Y.~Li, C.~Wu, Honeycomb lattice with multiorbital structure:
  Topological and quantum anomalous hall insulators with large gaps.
\newblock {\it Phys. Rev. B\/} {\bf 90}, 075114 (2014).

\bibitem{lee2010fwave}
W.-C. Lee, C.~Wu, S.~Das~Sarma, $f$-wave pairing of cold atoms in optical
  lattices.
\newblock {\it Phys. Rev. A\/} {\bf 82}, 053611 (2010).

\bibitem{wu2010f}
C.~Wu, Orbital ordering and frustration of $p$-band mott insulators.
\newblock {\it Phys. Rev. Lett.\/} {\bf 100}, 200406 (2008).

\bibitem{jacqmin2014}
T.~Jacqmin, I.~Carusotto, I.~Sagnes, M.~Abbarchi, D.~D. Solnyshkov,
  G.~Malpuech, E.~Galopin, A.~Lema\^{\i}tre, J.~Bloch, A.~Amo, Direct
  observation of dirac cones and a flatband in a honeycomb lattice for
  polaritons.
\newblock {\it Phys. Rev. Lett.\/} {\bf 112}, 116402 (2014).

\bibitem{liu2013omf}
Z.~Liu, Z.-F. Wang, J.-W. Mei, Y.-S. Wu, F.~Liu, Flat chern band in a
  two-dimensional organometallic framework.
\newblock {\it Phys. Rev. Lett.\/} {\bf 110}, 106804 (2013).

\bibitem{su2018}
N.~Su, W.~Jiang, Z.~Wang, F.~Liu, Prediction of large gap flat chern band in a
  two-dimensional metal-organic framework.
\newblock {\it Applied Physics Letters\/} {\bf 112}, 033301 (2018).

\bibitem{PhysRevB.90.094422}
A.~Smerald, F.~Mila, Exploring the spin-orbital ground state of
  ${\mathrm{ba}}_{3}{\mathrm{cusb}}_{2}{\mathrm{o}}_{9}$.
\newblock {\it Phys. Rev. B\/} {\bf 90}, 094422 (2014).

\bibitem{PhysRevB.98.245103}
J.~W.~F. Venderbos, R.~M. Fernandes, Correlations and electronic order in a
  two-orbital honeycomb lattice model for twisted bilayer graphene.
\newblock {\it Phys. Rev. B\/} {\bf 98}, 245103 (2018).

\bibitem{PhysRevB.100.205131}
W.~M.~H. Natori, R.~Nutakki, R.~G. Pereira, E.~C. Andrade, Su(4) heisenberg
  model on the honeycomb lattice with exchange-frustrated perturbations:
  Implications for twistronics and mott insulators.
\newblock {\it Phys. Rev. B\/} {\bf 100}, 205131 (2019).

\bibitem{kuc2011}
A.~Kuc, N.~Zibouche, T.~Heine, Influence of quantum confinement on the
  electronic structure of the transition metal sulfide $t$s${}_{2}$.
\newblock {\it Phys. Rev. B\/} {\bf 83}, 245213 (2011).

\bibitem{zhang2014direct}
Y.~Zhang, T.-R. Chang, B.~Zhou, Y.-T. Cui, H.~Yan, Z.~Liu, F.~Schmitt, J.~Lee,
  R.~Moore, Y.~Chen, {\it et~al.\/}, Direct observation of the transition from
  indirect to direct bandgap in atomically thin epitaxial mose 2.
\newblock {\it Nature nanotechnology\/} {\bf 9}, 111 (2014).

\bibitem{tokura2000orbital}
Y.~Tokura, N.~Nagaosa, Orbital physics in transition-metal oxides.
\newblock {\it science\/} {\bf 288}, 462--468 (2000).

\bibitem{corboz2012}
P.~Corboz, M.~Lajk\'o, A.~M. L\"auchli, K.~Penc, F.~Mila, Spin-orbital quantum
  liquid on the honeycomb lattice.
\newblock {\it Phys. Rev. X\/} {\bf 2}, 041013 (2012).

\bibitem{khomskii2003orbital}
D.~Khomskii, M.~Mostovoy, Orbital ordering and frustrations.
\newblock {\it Journal of Physics A: Mathematical and General\/} {\bf 36}, 9197
  (2003).

\bibitem{PhysRevB.90.075114}
G.-F. Zhang, Y.~Li, C.~Wu, Honeycomb lattice with multiorbital structure:
  Topological and quantum anomalous hall insulators with large gaps.
\newblock {\it Phys. Rev. B\/} {\bf 90}, 075114 (2014).

\bibitem{kresse93ab}
G.~Kresse, J.~Hafner, Ab initio molecular dynamics for liquid metals.
\newblock {\it Phys. Rev. B\/} {\bf 47}, 558 (1993).

\bibitem{blochl94}
P.~E. Bl{\"o}chl, Projector augmented-wave method.
\newblock {\it Phys. Rev. B\/} {\bf 50}, 17953 (1994).

\bibitem{pbe}
J.~P. Perdew, K.~Burke, M.~Ernzerhof, Generalized gradient approximation made
  simple.
\newblock {\it Phys. Rev. Lett.\/} {\bf 77}, 3865 (1996).

\bibitem{tsmethod09}
A.~Tkatchenko, M.~Scheffler, Accurate molecular van der waals interactions from
  ground-state electron density and free-atom reference data.
\newblock {\it Phys. Rev. Lett.\/} {\bf 102}, 073005 (2009).

\end{thebibliography}

\noindent \textbf{Acknowledgements:} 
This work is supported by the European Research Council (ERC-2015-AdG-694097), Grupos Consolidados (IT1249-19),  and SFB925. MC and AR are supported by the Flatiron Institute, a division of the Simons Foundation. We  acknowledge  funding 
by the Deutsche Forschungsgemeinschaft (DFG, German Research Foundation) under Germany's Excellence Strategy - Cluster  of  Excellence  Matter  and  Light  for Quantum Computing (ML4Q) EXC 2004/1 - 390534769 
and Advanced Imaging of Matter (AIM) EXC 2056 - 390715994 and funding by the Deutsche Forschungsgemeinschaft (DFG, German Research Foundation) under RTG 1995. Support by the Max Planck Institute - New York City  Center for Non-Equilibrium Quantum Phenomena is acknowledged.
DK, MMS, and ST acknowledge  support from the Deutsche Forschungsgemeinschaft (DFG, German Research Foundation), Projektnummer 277146847 -- CRC 1238 (projects C02, C03).

\end{document}